\documentclass[twocolumn,aps,prb,showpacs]{revtex4}
\usepackage{graphicx}
\usepackage{amssymb}

\def\ihp{\tilde{\varphi}}
\def\atr{\tilde{a}}

\begin{document}

\title{Semiclassical framework for the calculation of transport anisotropies}

\author{Karel V\'yborn\'y$^{1}$, Alexey A. Kovalev$^{2}$, 
Jairo Sinova$^{2,1}$, and T. Jungwirth$^{1,3}$}
\address{$^1\mbox{Institute of Physics,  Academy of Sciences of the Czech
  Rep., v.v.i., Cukrovarnick\'a 10, Praha 6 CZ--16253, Czech Republic}$}
\address{$^2$Department of Physics, Texas A\&M University, College Station, TX
77843-4242, USA}
\address{$^3$School of Physics and Astronomy, University of Nottingham, 
Nottingham NG7 2RD, United Kingdom}

\date{Oct 26th, 2008}



\begin{abstract}
We present a procedure for finding the exact solution to the linear-response 
Boltzmann equation for two-dimensional 
anisotropic systems and demonstrate it on examples 
of non-crystalline anisotropic
magnetoresistance in a system with spin-orbit interaction. 
We show that two decoupled integral
equations must be solved in order to find the non-equilibrium distribution
function up to linear order in the applied electric field. The examples
are all based on the 
Rashba system with charged magnetic scatterers, a system where 
the non-equilibrium distribution function and anisotropic magnetoresistance 
can be evaluated analytically.  Exact results are compared to earlier
widely-used approximative approaches. 
We find circumstances under which approximative approaches
may become unreliable even on a qualitative level.
\end{abstract}

\maketitle

\section{Introduction}

The change of electric resistance upon varying 
magnetization direction is an old
and well-known phenomenon\cite{Thomson:1857_a,Doring:1938_a,Jaoul:1977_a} with
applications in spintronics.\cite{Zutic:2004_a,Fabian:2007_a}  
Although the experimental
observation of this anisotropic magnetoresistance (AMR) is rather direct 
--- first accomplished as early as 1857 --- its theoretical
understanding is far from being complete.  
It has long been clear that the phenomenon arises from the combined effects of
magnetization and spin-orbit interaction. Disregarding the crystalline
anisotropic background, the magnetization-broken symmetry between two chosen 
directions and unequal resistivities
along these has been described within different models. In transition metal
ferromagnets, the anisotropy
was ascribed to asymmetric scattering due to different parts of the spin-orbit
interaction $\vec{L}\cdot\vec{S} = \frac{1}{2}(L_-S_+ + L_+ S_-) 
+ L_zS_z$ and the mechanisms considered were dubbed the
$L_-S_+$ model,\cite{Smit:1951_a} $L_zS_z$ model,\cite{Berger:1964_a} or a
combination of both.\cite{McGuire:1975_a,Jaoul:1977_a}   
Later, when computational power became sufficient for such task, 
ab initio calculations were performed\cite{Banhart:1995_a} for permalloy 
and reached a good agreement with experiments. 
However, no direct link between the ab initio 
and the model calculations listed above has been established, probably due to 
rather complex band structures involved.
On the other hand, such link between microscopic calculations and a simple
physical model was recently found in the diluted magnetic 
semiconductor\cite{Rushforth:2007_b,Rushforth:2007_a} (Ga,Mn)As
whose band structure is much simpler.

Despite the long history of the AMR research, the question has not been
answered to date of how a rigorous transport formalism for anisotropic
systems should be formulated. Instead, the transport anisotropy has often been
discussed only in terms of the asymmetry in scattering
amplitudes between two states on the Fermi surface. 
Current availability of materials with relatively simple band structure
motivates the quest for more precise theories of AMR.  The present article
discusses how the semiclassical Boltzmann equation should be solved in
anisotropic systems, using an example of the model two-dimensional (2D)
electron system. This allows us to put the previous approximations on 
rigorous grounds. 

The conductivity of a given solid in the regime of linear response to 
the electric field $\vec{\cal E}$ can be evaluated within the
semiclassical picture once we have found the distribution function satisfying
the Boltzmann equation. In the literature, this non-equilibrium
distribution function is approximated in several ways.
The relaxation time approximation (RTA) relies on calculating
the transport relaxation time $\tau$  from the scattering amplitudes
$w(\vec{k},\vec{k}')$ between two states on the Fermi surface using
\begin{equation}\label{eq-15}
  \frac{1}{\tau} =
  \int \frac{d^2 k'}{(2\pi)^2} w(\vec{k},\vec{k}') 
  \big[1-\cos\vartheta_{\vec{k}\vec{k}'}\big]\,.
\end{equation}
For isotropic systems, where $w$ depends only on the angle 
$\vartheta_{\vec{k}\vec{k}'}$ between
$\vec{k}$ and $\vec{k}'$, the integral~(\ref{eq-15}) does
not depend on the direction of $\vec{k}$ and the RTA provides in fact the
exact solution to the Boltzmann equation\cite{Ashcroft:1976_a}. 
The scattering rate $1/\tau$ depends
only on energy and it is constant on the whole Fermi surface once
the Fermi energy is fixed.

For anisotropic systems, keeping Eq.~(\ref{eq-15}) in use produces
$1/\tau$ that depends on the direction of $\vec{k}$. The 
non-equilibrium distribution function constructed 
utilising the RTA and Eq.~(\ref{eq-15})
may capture some aspects of the transport anisotropies but it is certainly not
precise.  This approximative approach underlies for example our previous
calculations\cite{Rushforth:2007_a} or those of McGuire and
Potter\cite{McGuire:1975_a} and we refer to it as to the ``$1/\tau$ approach''.

An improvement was proposed by Schliemann and 
Loss.\cite{Schliemann:2003_b} In what we will call
the ``$1/\tau^\parallel\,\&\,1/\tau^\perp$ approach'', they use, according to
their notation, Eq.~(\ref{eq-15})
to calculate $1/\tau^\parallel(\vec{k})$, and they
provide an explicit formula for the non-equilibrium distribution in terms of
this quantity and of
\begin{equation}\label{eq-16}
  \frac{1}{\tau^\perp(\vec{k})} =
  \int \frac{d^2 k'}{(2\pi)^2} w(\vec{k},\vec{k}') 
  \sin\vartheta_{\vec{k}\vec{k}'} \,.
\end{equation}

In our article, we argue that in a general case the non-equilibrium
distribution function cannot be exactly calculated by just evaluating two
integrals such as Eqs.~(\ref{eq-15},\ref{eq-16}) for each $\vec{k}$-point of
the Fermi surface separately. Instead, an integral
equation must be solved that determines the whole non-equilibrium distribution
at once. In Section~\ref{sec-II}, we describe this exact 
``integral equation approach'' to transport in anisotropic 
2D systems and then, in Section~\ref{sec-III}, we use a 
simple model system to illustrate how
the procedure works. For this purpose we introduce
the Rashba Hamiltonian combined with a scattering potential due to randomly
distributed charged and ferromagnetically ordered impurities. In this model,
the AMR results from the  spin-orbit coupled band structure and the broken
time-reversal symmetry of the scattering potential.\cite{Rushforth:2007_a}
We explicitly calculate exact non-equilibrium distribution functions for
several specific realizations of this model, starting from the ones with
simple solutions and then proceeding to the more complex case. 
Throughout Section~\ref{sec-III} we compare our distribution functions and
AMRs to results of the other two approximative approaches. 
Section~\ref{sec-IV} concludes the main body of the article
by discussing the relevance of our model calculations for the AMR in magnetic
semiconductors and by summarizing the key elements of the theoretical
framework we have developed.  The 
Appendices contain details of our calculations and also a more thorough
description of the $1/\tau$ and $1/\tau^\parallel\,\&\,1/\tau^\perp$
approaches.

\section{The framework}
\label{sec-II}

Our central goal is to obtain the distribution function 
$f=f(\vec{k},\vec{{\cal E}})$ of a conductor displaced from equilibrium
by a small bias represented by a weak 
homogeneous electric field $\vec{{\cal E}}$.
We start with the familiar form of the Boltzmann equation in 2D
\begin{equation}\label{eq-01}
  -e\vec{{\cal E}}\cdot\vec{v}(\vec{k})(-\partial_\epsilon f_0) =
  \int \frac{d^2 k'}{(2\pi)^2} 
w(\vec{k},\vec{k}')[f(\vec{k},\vec{\cal E})-f(\vec{k}',\vec{\cal E})]
\end{equation}
for a steady state of a spatially homogeneous system.
Derivation of this equation is described for instance in
Ref.~\onlinecite{Schliemann:2003_b}. Equation~(\ref{eq-01})  is valid up to
linear order in $|\vec{{\cal E}}|$ and it assumes small deviations of 
$f(\vec{k},\vec{E})$ from the equilibrium distribution $f_0=f_0(\vec{k})$. The 
velocity $\vec{v}=(1/\hbar)\nabla_{k} \epsilon_{\vec{k}}$ is implied by 
the band dispersion $\epsilon_{\vec{k}}$, and $e$ is the charge of carriers.
The scattering rate $w$ (per unit area of the reciprocal space) from the
state $\vec{k}$ to $\vec{k}'$ needs to
be specified according to the microscopic origin of the scattering; specific 
examples can be found in Appendix~\ref{sec-AppA} or in
Eq.~(38) of Ref.~\onlinecite{Schliemann:2003_b}. 
For now we only assume that the scattering is elastic, that is
$w(\vec{k},\vec{k}')\propto \delta\big(\epsilon_{\vec{k}}-
\epsilon_{\vec{k}'}\big)$. Focusing on the AMR, we do not include
anomalous terms\cite{Sinitsyn:2006_b,Sinitsyn:2007_a} like the
coordinate shift related to the side jump in the anomalous Hall effect
into the right hand side of Eq.~(\ref{eq-01}).
Equation~(\ref{eq-01}) can be applied to multi-band systems where
$\vec{k}$ is replaced by a compound index containing the wavevector
and band index and the integral by integration over the wavevector and
summation over bands.

The solution to Eq.~(\ref{eq-01}) is a function both of $\vec{k}$ and
$\vec{{\cal E}}$. Focusing first on the latter, we can write it as
Taylor series 
\begin{equation}\label{eq-02}
f(\vec{k},\vec{{\cal E}}) =
  f_{0}+{\cal E}_x \partial_{{\cal E}_x}f+{\cal E}_y \partial_{{\cal E}_y} f+
\sum_{ij}{\cal E}_i{\cal E}_j\partial_{{\cal E}_i}\partial_{{\cal E}_j} f +
\ldots
\end{equation}
Being interested only in the linear order of the electric field components
$({\cal E}_x,{\cal E}_y)$, we will truncate the series after the third term.
For the simplicity of notation, we will now assume that the band structure is
isotropic, $\epsilon_{\vec{k}}=\epsilon_k$ and $-\delta_\epsilon
f_0=\delta(\epsilon-\epsilon_k)$. The anisotropy can still pervade into the
transport via $w$, for instance due to anisotropic scatterers. The more
general Boltzmann equation for anisotropic bands is treated in
Appendix~\ref{sec-AppG}. 

We define two
angles $\phi$, $\theta$ as $\vec{{\cal E}}={\cal E}(\cos\theta,\sin\theta)$
and $\vec{k}=k(\cos\phi,\sin\phi)$ and Eq.~(\ref{eq-02}) becomes
\begin{equation}\label{eq-03}
f(\phi,\theta) -f_0 = 
         {\cal E}\big(A(\phi)\cos\theta + B(\phi)\sin\theta\big)
\end{equation}
after the truncation, where $A(\phi)=\partial_{{\cal E}_x}f$ and
$B(\phi)=\partial_{{\cal E}_y}f$. 
The non-equilibrium distribution is now expressed in terms of two functions of
$\phi$ which must, according to Eq.~(\ref{eq-01}) with Eq.~(\ref{eq-03})
inserted, fulfil
\begin{eqnarray} \nonumber
  \cos(\theta-\phi) &=&
\left[\bar{w}(\phi)a(\phi) -
            \int d\phi'\ w(\phi,\phi')a(\phi') \right]\cos\theta +\\
 && \label{eq-04} \hskip-1cm
+\left[\bar{w}(\phi)b(\phi) -
            \int d\phi'\ w(\phi,\phi')b(\phi') \right]\sin\theta\,.
\end{eqnarray}
We define here $A(\phi) \equiv -ev(-\partial_\epsilon f_0)\, a(\phi)$,
$B(\phi) \equiv -ev(-\partial_\epsilon f_0)\, b(\phi)$,
and $\bar{w}(\phi)=\int d\phi'\ w(\phi,\phi')$, where
$w(\phi,\phi')=(2\pi)^{-2}\int k'\, dk' w(\vec{k},\vec{k}')$ now includes the
original transport scattering rate $w(\vec{k},\vec{k}')$ 
and also the density of states.

The integral equation~(\ref{eq-04}) with two variables $\phi,\theta$ can be
decomposed into two independent integral equations
\begin{eqnarray}\label{eq-05a}
  \cos\phi &=& \bar{w}(\phi)\, a(\phi) - \int d\phi' w(\phi,\phi') a(\phi')
\\[-1mm]
\label{eq-05b}
  \sin\phi &=& \bar{w}(\phi)\, b(\phi) - \int d\phi' w(\phi,\phi') b(\phi')
\end{eqnarray}
whose solutions $a(\phi)$, $b(\phi)$ 
inserted into Eq.~(\ref{eq-03}) yield the exact
solution of the Boltzmann equation~(\ref{eq-01}) up to the linear order in
${\cal E}$.

The two decoupled inhomogeneous Fredholm equations\cite{fredholm} of the second
kind~(\ref{eq-05a},\ref{eq-05b}) can be most conveniently solved in terms of
Fourier series. For special choices of $w(\phi,\phi')$ or band structure
anisotropy (see Appendix~\ref{sec-AppG}), 
the series may contain only few terms and reduce to an ansatz for
$f(\phi,\theta)$ such as Eq.~(15) in Ref.~\onlinecite{Trushin:2006_a}. We
explain the general procedure how to solve Eqs.~(\ref{eq-05a},\ref{eq-05b}) on
three examples below.

\section{Three examples with Rashba system}
\label{sec-III}

To illustrate how the above outlined procedure works, we choose the 
$2\times 2$ Rashba Hamiltonian\cite{Fabian:2007_a} in the basis of plane
waves 
\begin{equation}\label{eq-06}
  \hat{H} = \frac{\hbar^2 k^2}{2m} + \lambda(k_x\sigma_y - k_y\sigma_x)\,,
\end{equation}
where $\sigma_{x,y}$ are the Pauli matrices, and $\lambda$ is the Rashba
parameter. In addition to Eq.~(\ref{eq-06}), we consider scattering on
dilute charged magnetic impurities\cite{Rushforth:2007_a,Rushforth:2007_b} 
described by the operator $\hat{V}$,
\begin{equation}\label{eq-07}
  \hat{V}/V_0 = \alpha + \sigma_x = 
  \left(\begin{array}{cc}\alpha & 1 \\ 1 & \alpha \end{array}\right)\,,
\end{equation}
that is impurities containing short range electric and ferromagnetically
ordered magnetic potentials.
The quantity $\alpha$ is the (dimensionless) strength of the electric part,
relative to the magnetic part, of the 'electro-magnetic scatterer' whose
magnetic moment was chosen to be along the $x$
direction. The magnitude $V_0$ and other aspects of this model are
discussed in Section~\ref{sec-IV} and Appendices~\ref{sec-AppA}
and~\ref{sec-AppB}.

We now calculate the non-equilibrium Boltzmann distribution function
$f(\vec{k},\vec{\cal E})=f(\phi,\theta)$ for this
model in several special cases. To facilitate relevant comparison between the
$1/\tau$ and $1/\tau^\parallel\,\&\,1/\tau^\perp$ approaches
and the exact integral equation approach of
Section~\ref{sec-II}, we calculate $f(\phi,\theta)$ and 
evaluate the AMR within the approximative approaches as well.

\subsection{Single band and magnetic scatterers}

\begin{figure}
\includegraphics[scale=1.6]{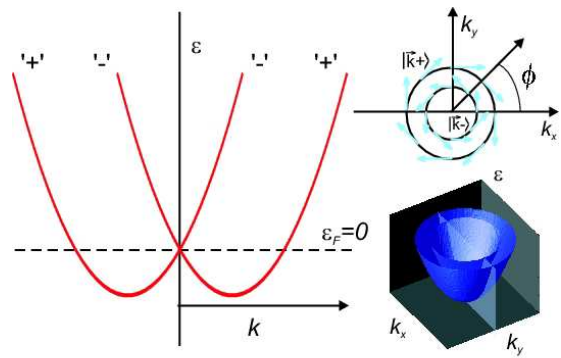}
\caption{Summary of the Rashba model defined by the
  Hamiltonian~(\ref{eq-06}). The three-dimensional 
  plot shows energy dispersions; its cross section along one (arbitrary)
  direction in the $\vec{k}$-space is also shown. The spin textures 
  for the pair of eigenstates $|\vec{k}+\rangle$ and $|\vec{k}-\rangle$ for
  each $\vec{k}$ at the respective Fermi surfaces 
  with $\epsilon_F>0$ are shown on the
  top. Unlike the diameters of the $+$ and $-$ Fermi surfaces, 
  the eigenstates (spin textures) are independent of the Fermi energy
  $\epsilon_F$, according to Eq.~(\ref{eq-35}).}
\label{fig-02}
\end{figure}

The first special case of the model above concerns purely magnetic scatterers
($\alpha=0$) in the situation when the Fermi energy cuts the spectrum of the
Rashba Hamiltonian~(\ref{eq-06}) precisely at the $k=0$ degeneracy point
($\epsilon_F=0$ in Fig.~\ref{fig-02}). We further disregard this single point
of the Fermi surface and consider only the '+' band.  This case offers the
simplest way to explain the calculation of $f$ outlined in
Section~\ref{sec-II}.

The dimensionless scattering probability corresponding
to $\hat{V}$ of~Eq.~(\ref{eq-07}) is 
\begin{equation}\label{eq-08}
  w(\phi,\phi')/K = \frac{1}{2}[1-\cos(\phi+\phi')]\,.
\end{equation}
This result, including the dimensionful prefactor $K$, is derived using the
Fermi golden rule in Appendix~\ref{sec-AppA}~and~\ref{sec-AppB}. 
Although $w(\phi,\phi')$ does not explicitly depend on the Rashba
parameter $\lambda$, the presence of the spin-orbit coupling, combined with the
symmetry breaking scattering potential, has the crucial implication that $w$
depends on absolute values of  angles $\phi$ and $\phi'$. This leads
to the non-zero anisotropy of the magnetotransport, in contrast to  the
isotropic case in which $w(\phi,\phi')$ depends only on the relative
angle $\phi-\phi'$ between the incoming and outgoing momenta. 
The total scattering probability, implied by Eq.~(\ref{eq-08}), reads
\begin{equation}\label{eq-09}
\bar{w}(\phi) = K\pi\,.
\end{equation}
Note that despite the
independence of $\bar{w}$ on $\phi$ in the special case considered in this
subsection the resulting relaxation times and conductivity are indeed
anisotropic.


We will now look for the solution $a(\phi)$ to Eq.~(\ref{eq-05a}) in the form
of Fourier series
\begin{eqnarray}\nonumber
  a(\phi) &=& a_{0} + a_{c1}\cos\phi + a_{c2}\cos 2\phi + \ldots +\\
  \label{eq-10}
  && \phantom{a_{0}} + a_{s1}\sin\phi + a_{s2}\sin 2\phi + \ldots\,.
\end{eqnarray}
%
Owing to the trivial form of $w(\phi,\phi')$ (of its Fourier
spectrum) the integral in Eq.~(\ref{eq-05a}) can be readily 
calculated and Eq.~(\ref{eq-05a}) assumes the following form:
\begin{eqnarray*}\displaystyle
  \frac{1}{\pi K}\cos\phi &=& \frac{3}{2}a_{c1}\cos\phi + a_{c2}\cos 2\phi +
  a_{c3}\cos 3\phi+ \ldots + \\
                             &&  + \frac{1}{2}a_{s1}\sin\phi+
                             a_{s2}\sin 2\phi + a_{s3}\sin 3\phi + \ldots\,.
\end{eqnarray*}
%
The only non-zero coefficients in the Fourier series~(\ref{eq-10}) are
therefore $a_0$ and $a_{c1}$. The solution of Eq.~(\ref{eq-05a}) then reads
\begin{equation}\label{eq-12}
  a(\phi) = a_0 + \frac{2/3}{\pi K}\cos\phi\,.
\end{equation}
Conservation of the number of particles requires $a_0$ to be zero.

A completely analogous procedure applied to 
Eq.~(\ref{eq-05b}) yields a system of equations for coefficients $b_0$,
$b_{s1}$, $b_{c1}$ which give
%
\begin{equation}\label{eq-13}
  b(\phi) = \frac{2}{\pi K}\sin\phi\,.
\end{equation}
The complete solution up to linear order in ${\cal E}$ to the Boltzmann
equation~(\ref{eq-01}) written using Eq.~(\ref{eq-03}) is therefore
\begin{equation}\label{eq-14}
  f(\phi,\theta)\!=\!f_0 - ev{\cal E}(-\partial_\epsilon f_0) \frac{2}{\pi K}
  \bigg[\frac{1}{3}\cos\theta\cos\phi + \sin\theta\sin\phi\bigg]\,.
\end{equation}

Let us now compare this result to the approximate approaches outlined in the
Introduction. The non-equilibrium distribution in the $1/\tau$ approach is (see
Appendix~\ref{sec-AppC}) 
\begin{eqnarray}\label{eq-17}
  f(\phi,\theta) - f_0 &=& - ev{\cal E}(-\partial_\epsilon f_0) \frac{2}{\pi K}
  \times
  \\ \nonumber
  &&\hskip-2mm\times\left[\cos\theta\frac{\cos\phi}{3-2\sin^2\phi}
          + \sin\theta\frac{\sin\phi}{3-2\sin^2\phi}\right]\,,
\end{eqnarray}
while in the $1/\tau^\parallel\, \&\, 1/\tau^\perp$ approach (see
Appendix~\ref{sec-AppD}), we obtain
\begin{eqnarray}\label{eq-18}
  f(\phi,\theta) - f_0 &=& - ev{\cal E}(-\partial_\epsilon f_0) \frac{2}{\pi K}
  \times
  \\ \nonumber
  && \times\left[\cos\theta
  \frac{3\cos\phi+2\sin^2\phi(1-\cos\phi)}{9+4\sin^4\phi-8\sin^2\phi} + 
  \right.\\
  && \hskip.7cm \nonumber \left.
  +\sin\theta
  \frac{3\sin\phi(1-\cos\phi)-2\sin^3\phi}{9+4\sin^4\phi-8\sin^2\phi}\right]\,.
\end{eqnarray}
Distribution functions
in Eqs.~(\ref{eq-14},\ref{eq-17},\ref{eq-18}) are significantly different.
To quantify the differences, we use these three distribution functions
to calculate the AMR, defined as
\begin{equation}\label{eq-19}
  \mbox{AMR} = -\frac{\sigma_{xx}-\sigma_{yy}}{\sigma_{xx}+\sigma_{yy}}
\end{equation}
and having the meaning of the (relative) difference in resistivity for current
flowing parallel and perpendicular to the direction of the scatterer's
magnetic moment, respectively. The conductivities are calculated from the
current implied by the non-equilibrium distribution $f(\vec{k},\vec{\cal E})$
\begin{equation}\label{eq-57}
  \vec{j}(\vec{\cal E}) = \int \frac{d^2 k}{(2\pi)^2} e\vec{v}(\vec{k})\,
  f(\vec{k},\vec{\cal E})\,,
\end{equation}
i.e., as
$\sigma_{xx}=j(\theta=0)/{\cal E}$ and $\sigma_{yy}=j(\theta=\pi/2)/{\cal E}$
where $j(\theta)=\int d\phi f(\phi,\theta)|\vec{v}|\cos(\phi-\theta)$.

The AMR value of $1/2$, obtained from the exact non-equilibrium
distribution function in Eq.~(\ref{eq-14}), is markedly different from the
results of the approximative approaches. The $1/\tau$ approach underestimates
the AMR by almost a factor of two ($\mbox{AMR}\approx 0.27$), and the 
$1/\tau^\parallel\, \&\, 1/\tau^\perp$ approach 
predicts even a wrong sign ($\mbox{AMR}\approx -0.11$).

Before we proceed to comparing the three approaches on other realizations of
our model disordered 2D system,
let us make a remark about the distribution functions above. The
non-equilibrium part of the
distribution function in Eq.~(\ref{eq-17}) was obtained as 
$- e\vec{v}\cdot{\cal \vec{E}}(-\partial_\epsilon f_0) \tau(\phi)$ 
with $\vec{v}\cdot\vec{\cal E}=v{\cal E}\cos(\theta-\phi)$ and 
$\tau(\phi)=(2/\pi K)(3-2\sin^2\phi)^{-1}$ as derived in
Appendix~\ref{sec-AppC}. 
Analogous factorization of the bracket in Eq.~(\ref{eq-18}) or
Eq.~(\ref{eq-14}) is not possible, reflecting the fact that no scalar
relaxation time can be attributed to a given $\vec{k}$--state in these
approaches. However, the $1/\tau^\parallel\,\&\, 1/\tau^\perp$ approach
still unambiguously assigns relaxation-rate-like quantities, a pair of (not
necessarily positive) values
$1/\tau^\parallel(\vec{k}), 1/\tau^\perp(\vec{k})$, to each $\vec{k}$-state,
independent of the electric field direction (determined by $\theta$; see
Appendix~\ref{sec-AppD}). It remains an open question whether also the 
exact solution of the Boltzmann equation, such as Eq.~(\ref{eq-14}), can be
meaningfully interpreted in terms of $\theta$--independent quantities related
to scattering.

\subsection{Single band and electro-magnetic scatterers}

We now extend results of the previous section by relaxing the condition
$\alpha=0$, that is we consider the complete scatterer with electric and
magnetic parts of its potential added up coherently, as defined by
Eq.~(\ref{eq-07}).  The extension is straightforward although the algebra
involved is richer than for the previous model.  The dimensionless scattering
probability $w(\phi,\phi')/K$ and $\bar{w}(\phi)$ are
\begin{equation}\label{eq-20}
  \begin{array}{rcl}
  w(\phi,\phi')/K &=& \frac{1}{2}\big[1-\cos(\phi+\phi')+
     \alpha^2\big(1+\cos(\phi-\phi')\big)\big] \\[1mm]
   && + \alpha(\sin\phi + \sin\phi') \\[2mm]
  \bar{w}(\phi) &=& \pi K(1+\alpha^2+2\alpha\sin\phi)\,,
  \end{array}
\end{equation}
%
as shown in Appendices~\ref{sec-AppA}~and~\ref{sec-AppB}. Note that 
$\bar{w}(\phi)$ is no longer constant, which is here the direct reason of
the more complex algebra needed. We again look for the solution of
Eq.~(\ref{eq-05a}) in the form of Fourier series~(\ref{eq-10}) and find that
the higher order coefficients $a_{c2},a_{c3},\ldots$ are now no longer zero.
Instead of
Eq.~(\ref{eq-12}), we get a system of an infinite number of linear equations
which  is not surprising, given that Eq.~(\ref{eq-05a}) is an
integral equation in its general form.

This system of equations can be solved using a partitioning method, described
in Appendix~\ref{sec-AppE}. 
Herein, we segregate the variables into three groups:
$\{a_0,a_{c1},a_{s1}\}$, $\{a_{c2},a_{c3},\ldots\}$, and
$\{a_{s2},a_{s3},\ldots\}$. The first group must obey
\begin{equation} 
 \label{eq-21}\begin{array}{rcl}
 1/(\pi K)  &=& 
    (1+\alpha^2)a_{c1} + \alpha a_{s2} + \frac{1}{2}(1-\alpha^2)a_{c1} \\[1mm]
 0 &=& (1+\alpha^2)a_{s1} - \alpha a_{c2} - \frac{1}{2}a_{s1}(1+\alpha^2)\,,
\end{array}
\end{equation}
and $a_0=0$ as in the previous Subsection. Equations~(\ref{eq-21}) 
originate from comparing the coefficients in front of the
$\cos\phi$ and $\sin\phi$ terms of Eq.~(\ref{eq-05a}) with
Eq.~(\ref{eq-10}) inserted. Separate treatment of the other two infinite
systems of equations, described in Appendix~\ref{sec-AppE}, yields
\begin{equation}\label{eq-22}
 a_{c2}=
\left\{\begin{array}{l} a_{s1}\alpha\\a_{s1}/\alpha \end{array}\right. \quad
 a_{s2}=\left\{\begin{array}{ll} -a_{c1}\alpha&\mbox{ for } |\alpha|< 1\\
                                 -a_{c1}/\alpha&\mbox{ for }|\alpha|> 1\,.
                               \end{array}\right.
 \end{equation}
Together, Eqs.~(\ref{eq-21},\ref{eq-22}) form a closed system 
for $a_{c1}$, and $a_{s1}$ which thus read 
$$
  a_{c1} = \frac{1}{2\pi K}\times \left\{\begin{array}{ll} 
4/(3-\alpha^2) & \mbox{ for }|\alpha|< 1\\ 4/(1+\alpha^2) & 
                                           \mbox{ for }|\alpha|> 1 
                                           \end{array}\right.\quad
  a_{s1} = 0\,.
$$
The solution to Eq.~(\ref{eq-05a}) for $|\alpha|< 1$ is then
\begin{equation}\label{eq-23}
  a(\phi) = \frac{1}{2\pi K}\cdot \frac{4}{3-\alpha^2}\cos\phi + 
            a_{c2}\cos 2\phi + a_{s2}\sin 2\phi + \ldots\,.
\end{equation}
For the evaluation of current and AMR using Eq.~(\ref{eq-19}) 
there is no need to know the higher order terms of $f(\phi,\theta)$
(by virtue of $\int_0^{2\pi} \cos\phi \cos 2\phi =0$ etc.).
However, keeping {\em all} higher order terms of~Eq.~(\ref{eq-10}) in the 
derivation was necessary for obtaining the xscorrect form of Eq.~(\ref{eq-21})
and also correct expressions for constants $a_{c1}$ and $a_{s1}$ at the end.

We again repeat the same procedure for Eq.~(\ref{eq-05b}), obtain $b(\phi)$,
and finally we complete the calculation by writing down the non-equilibrium
distribution function:
\begin{eqnarray}\label{eq-24a}
  f(\phi,\theta) - f_0 &=& - ev{\cal E}(-\partial_\epsilon f_0) \frac{2}{\pi K}
  \times
  \\
  \nonumber
  &&\times \left[\cos\theta \frac{\cos\phi}{3-\alpha^2} +\ldots +
        \sin\theta \frac{\sin\phi}{1-\alpha^2} +\ldots \right]
\end{eqnarray}
for $|\alpha|<1$, while for $|\alpha|>1$ the bracket is replaced by
\begin{equation}\label{eq-24b}
  \left[\cos\theta \frac{\cos\phi}{\alpha^2+1} +\ldots +
        \sin\theta \frac{\sin\phi}{\alpha^2-1} +\ldots \right]\,.
\end{equation}
The dots symbolize $\cos 2\phi$, $\sin 2\phi$ and higher order terms which as
emphasised above do not contribute to the AMR. The 
divergence of this expression for $|\alpha|\to 1$ will be discussed in
Section~\ref{sec-IV}. 

Evaluating the AMR using the distribution function~(\ref{eq-24a},\ref{eq-24b})
and Eqs.~(\ref{eq-57},\ref{eq-19}) amounts to comparing the coefficients in
front of the $\cos\theta\cos\phi$ and $\sin\theta\sin\phi$ summands. We get
\begin{equation}\label{eq-25}
  \mbox{AMR} = \frac{1}{2-\alpha^2}\,,\ |\alpha|<1\qquad 
  \mbox{AMR} = \frac{1}{\alpha^2}\,,\ |\alpha|>1\,.
\end{equation}
%

\begin{figure}
\begin{center}
\begin{tabular}{cc}
\hskip-1.8cm(a) & \hskip-4cm(b) \\
\hskip-.5cm\includegraphics[scale=0.57]{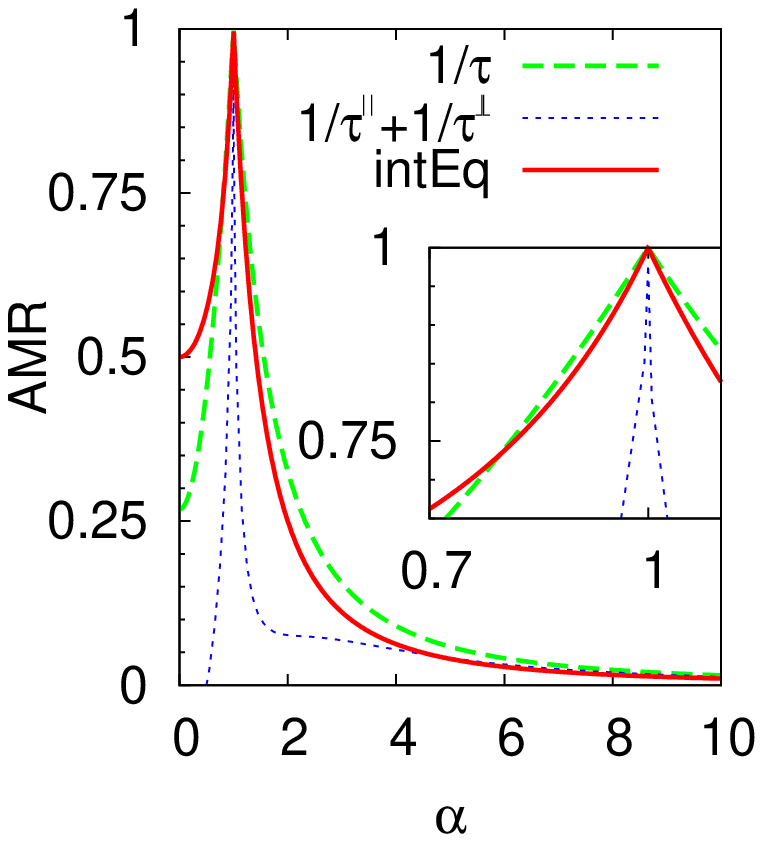} &
\hskip-3cm\includegraphics[scale=0.57]{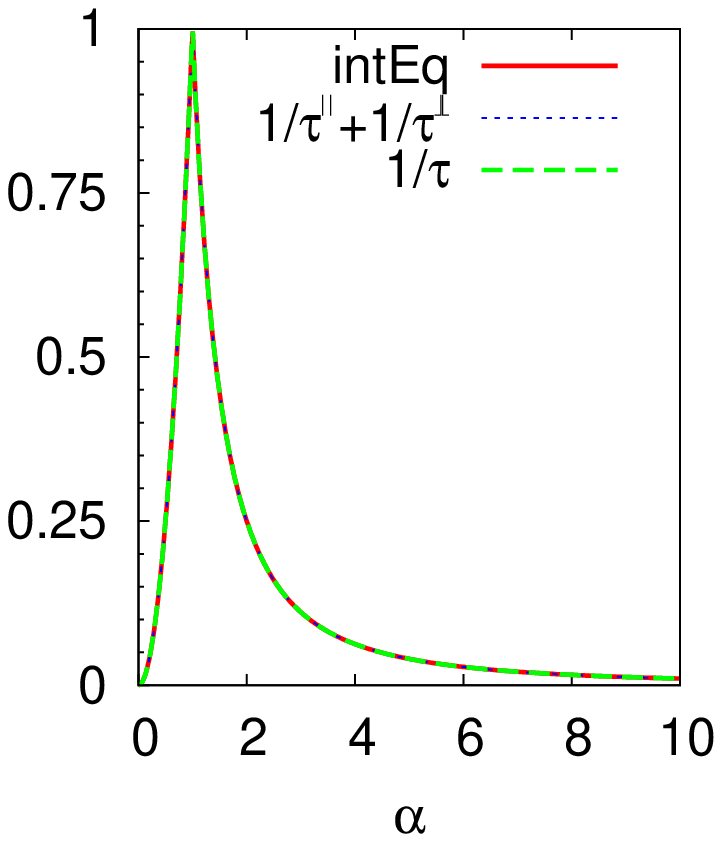}
\end{tabular}
\end{center}
\caption{AMR as a function of the relative strength, $\alpha$,  of the
  electric and magnetic components of the impurity potential 
  (see Eq.~(\ref{eq-07}) for the definition of $\alpha$).  
Dashed and dotted lines denote results of 
the approximative $1/\tau$ and $1/\tau^\parallel\,\&\,1/\tau^\perp$ 
approaches, solid line shows the exact result of the integral equation
approach. (a) Single-band model, (b) two-band model with 
$\lambda\to 0$. }
\label{fig-01}
\end{figure}

We conclude the study of the single-band model by comparing this AMR to the
results of the approximate 
$1/\tau$ and $1/\tau^\parallel\,\&\,1/\tau^\perp$ approaches
shown in Fig.~\ref{fig-01}(a). While the $1/\tau$ approach can be regarded as
only quantitatively inaccurate, as already suggested by the results of the
previous Subsection, the apparently more sophisticated
$1/\tau^{\parallel}\,\&\,1/\tau^{\perp}$ approach yields remarkably large
deviations from the exact AMR. 

\subsection{Two bands and electro-magnetic scatterers}

We now consider the case when the Fermi energy is above the $k=0$ degeneracy
point of the Rashba bands. 
Let us first explicitly write down the scheme of Sec.~\ref{sec-II} for a
two-band system. Considering distribution functions of the '+'
and '$-$' bands, denoted by $f_+$ and $f_-$, 
Eq.~(\ref{eq-01}) is replaced by two coupled equations
\begin{eqnarray}\label{eq-26}
  -e\vec{{\cal E}}\cdot\vec{v}_+(\vec{k})
\left(
 -\frac{\partial f_0(\epsilon_{\vec{k}+})}{\partial\epsilon}\right) 
  &=& \\ \nonumber
  && \hskip-4cm
  =\int \frac{d^2 k'}{(2\pi)^2} 
 \big[\big(w_{++}(\vec{k},\vec{k}') +w_{+-}(\vec{k},\vec{k}')\big)f_+(\vec{k})-
  \\ \nonumber
  && \hskip-2.5cm - w_{++}(\vec{k},\vec{k}')f_+(\vec{k}') 
  - w_{+-}(\vec{k},\vec{k}')f_-(\vec{k}')\big]
\end{eqnarray}
and
\begin{eqnarray*}
  -e\vec{{\cal E}}\cdot\vec{v}_-(\vec{k})
\left(
 -\frac{\partial f_0(\epsilon_{\vec{k}-})}{\partial\epsilon}\right) 
  &=& \\ \nonumber
  && \hskip-4cm
  =\int \frac{d^2 k'}{(2\pi)^2} 
 \big[\big(w_{--}(\vec{k},\vec{k}') +w_{-+}(\vec{k},\vec{k}')\big)f_-(\vec{k})-
  \\ \nonumber
  && \hskip-2.5cm - w_{--}(\vec{k},\vec{k}')f_-(\vec{k}') 
  - w_{-+}(\vec{k},\vec{k}')f_+(\vec{k}')\big]\,.
\end{eqnarray*}
The scattering rate $w_{fi}$ now also bears the indices of the initial ($i$)
and final ($f$) band. We will abbreviate the equilibrium distributions
$f_0(\epsilon_{\vec{k}\pm})$ by $f_{0\pm}$.  

Assuming isotropic bands $\epsilon_{k+}$ and $\epsilon_{k-}$, 
we seek a solution of Eqs.~(\ref{eq-26}) in the form of
\begin{equation}\label{eq-27}
\begin{array}{c}
\!\! f_+(\phi,\theta)\! -\! f_0 \!= \!
     -e{\cal E}v_+ (-\partial_\epsilon f_{0+})
       \big(a_+(\phi)\cos\theta + b_+(\phi)\sin\theta\big)\\[2mm]
\!\! f_-(\phi,\theta)\! -\! f_0 \!=\! 
     -e{\cal E}v_- (-\partial_\epsilon f_{0-})
       \big(a_-(\phi)\cos\theta + b_-(\phi)\sin\theta\big)
\end{array}
\end{equation}
and the four functions $a_\pm(\phi)$, $b_\pm(\phi)$ must fulfil
\begin{eqnarray}\label{eq-28a}
  \hskip-5mm&\begin{array}{rcl}
  \cos\phi &=& \bar{w}_+(\phi)\, a_+(\phi) \\
  &&\displaystyle  \hskip-5mm
  -\int d\phi'\left[ 
w_{++}(\phi,\phi') a_+(\phi') + w_{+-}(\phi,\phi') a_-(\phi')\right]\\
  \cos\phi &=& \bar{w}_-(\phi)\, a_-(\phi) \\
  &&\displaystyle  \hskip-5mm
  -\int d\phi' 
\left[w_{--}(\phi,\phi') a_-(\phi') + w_{-+}(\phi,\phi') a_+(\phi')\right]
  \end{array}& \\[5mm]
  \label{eq-28b}
  \hskip-5mm&\begin{array}{rcl}
  \sin\phi &=& \bar{w}_+(\phi)\, b_+(\phi) \\
  &&\displaystyle    \hskip-5mm
  -\int d\phi' 
\left[w_{++}(\phi,\phi') b_+(\phi') + w_{+-}(\phi,\phi') b_-(\phi')\right]\\
  \sin\phi &=& \bar{w}_-(\phi)\, b_-(\phi) \\
  &&\displaystyle    \hskip-5mm
  -\int d\phi'
\left[w_{--}(\phi,\phi') b_-(\phi') + w_{-+}(\phi,\phi') b_+(\phi')\right]
  \end{array}& 
\end{eqnarray}
where $w_{fi}(\phi,\phi')=(2\pi)^{-2} 
\int k'\, dk' w_{fi}(\vec{k},\vec{k}')$ and
$\bar{w}_{i}(\phi) = \int d\phi' [w_{+i}(\phi,\phi')+w_{-i}(\phi,\phi')]$.
Note that Eqs.~(\ref{eq-28a}) are decoupled from Eqs.~(\ref{eq-28b}).

\subsubsection{Evaluation of $f_+$ and $f_-$}

The dimensionless scattering probabilities for the complete electro-magnetic 
scattering operator given by Eq.~(\ref{eq-07}) are
\begin{equation}\label{eq-29}
  \begin{array}{rcl}
    w_{(++/--)}(\phi',\phi)/K &=& \\ &&\hskip-2.3cm =
     \frac{1}{2}\big[1-\cos(\phi+\phi')
     +\alpha^2\big(1+\cos(\phi-\phi')\big)\big]+\\
     &&\hskip1.5cm (+/-) \alpha(\sin\phi + \sin\phi')\\[3mm]
    w_{-+}(\phi',\phi)/K &=& w_{+-}(\phi,\phi')/K = \\ &&\hskip-2cm  =
     \frac{1}{2}\big[1+\cos(\phi+\phi')
     +\alpha^2\big(1-\cos(\phi-\phi')\big)\big]+\\
     &&\hskip2.2cm +\alpha(\sin\phi -\sin\phi')
  \end{array}
\end{equation}
and $\bar{w}_\pm(\phi)=2\pi K(1+\alpha^2\pm 2\alpha\sin\phi)$. 
For simplicity, we assume
that the constant $K$ (and the density of states, as explained in
Appendix~\ref{sec-AppG}) is the same for both bands. 
This occurs in the Rashba model~(\ref{eq-06}) in the limit of
$\lambda k_F\ll \epsilon_F$ and we will call this the $\lambda\to 0$
limit. Details of the derivation of Eq.~(\ref{eq-29}) are given in
Appendix~\ref{sec-AppA}~and~\ref{sec-AppB}. 

In a close analogy to the single-band case, equations~(\ref{eq-28a}) produce
two coupled infinite sets of linear equations for coefficients of
\begin{eqnarray}\nonumber
  a_\pm(\phi) &=& a_{0\pm}+ a_{c1\pm}\cos\phi + a_{c2\pm}\cos 2\phi + \ldots+\\
  \label{eq-30}
  && \phantom{a_{0}} + a_{s1\pm}\sin\phi + a_{s2\pm}\sin 2\phi + \ldots\,.
\end{eqnarray}
These may again be reduced to two coupled $3\times 3$ systems 
for variables $a_{0\pm},a_{c1\pm},a_{s1\pm}$
using the partitioning method.
Mathematically, their solution
\begin{eqnarray}\nonumber
  a_{0+} - a_{0-}   &=& 0  
  \\ \nonumber
  a_{c1+} = a_{c1-} &=& \displaystyle 
          \frac{1}{\alpha}\cdot \frac{1}{2\pi K}\times\left\{
  \begin{array}{ll}\alpha &\mbox{ for }|\alpha|\le 1\\
                   1/\alpha&\mbox{ for }|\alpha|\ge 1
    \end{array}\right.\\ \label{eq-52}
  a_{s1+} = a_{s1-} &=& 0 
\end{eqnarray}
leaves $a_{0+}+a_{0-}$ undetermined and, physically, 
particle number conservation again dictates that this constant is zero.
Terms in Eq.~(\ref{eq-30})
containing higher multiples of $\phi$ are again not contributing to the
current and to the AMR but their coefficients can be evaluated within the
partitioning procedure.

Applying the same procedure to Eqs.~(\ref{eq-28b}) leads to
\begin{eqnarray*}\nonumber
 b_\pm(\phi) &=& b_{0\pm} + b_{c1\pm}\cos\phi + b_{c2\pm}\cos 2\phi + \ldots+\\
  && \phantom{a_{0}} + b_{s1\pm}\sin\phi + b_{s2\pm}\sin 2\phi + \ldots
\end{eqnarray*}
with
\begin{eqnarray*}\nonumber
  b_{0+} - b_{0-}   &=& \frac{-2\alpha}{1+\alpha^2}b_s \\ \nonumber
  b_{c1+} = b_{c1-} &=& 0 \\
  b_{s1+} = b_{s1-} &=& b_s = \displaystyle 
          \frac{\alpha^2+1}{\alpha^2-1}\cdot\frac{1}{2\pi K}\times\left\{
  \begin{array}{ll}(-1)&\mbox{ for }|\alpha|\le 1\\
                   1/\alpha^2&\mbox{ for }|\alpha|\ge 1
    \end{array}\right.\\ 
\end{eqnarray*}
The non-zero value of $b_{0+} - b_{0-}$ means that the scattering
redistributes particles between the two bands. In another system, where the
two bands would have different net spin polarization, such redistribution would
correspond to the polarization of the particles by impurities. The overall
particle number conservation nevertheless again requires $b_{0+}+b_{0-}=0$.

The two non-equilibrium distribution functions are now
\begin{eqnarray}\label{eq-31}
  f_\pm(\phi,\theta) &=& f_0 - e{\cal E}v 
  (-\partial_\epsilon f_{0\pm}) \displaystyle\frac{1}{2\pi K} 
  \times \\
  \nonumber && \hskip-2cm 
  \left[\cos\theta\cos\phi + \ldots + 
        \frac{1+\alpha^2}{|1-\alpha^2|}\sin\theta\sin\phi + \ldots 
        \pm \frac{-\alpha}{|1-\alpha^2|}\sin\theta \right] 
\end{eqnarray}
for $|\alpha|\le 1$. The distribution function 
for $|\alpha|>1$ is given by Eq.~(\ref{eq-31}) with the term in the square
brackets multiplied by $1/\alpha^2$.

\subsubsection{Comparison to the approximate approaches}

The explicit calculation outlined in Appendix~\ref{sec-AppF} shows 
that again all
coefficients appearing in front of the cosine terms of Eq.~(\ref{eq-30}) are
non-zero. The infinite series, however, can be summed up and the complete exact
non-equilibrium distributions fulfilling Eqs.~(\ref{eq-26}) read
\begin{eqnarray}\label{eq-32}
  f_\pm(\phi,\theta) &=& f_0 - e{\cal E}v 
  (-\partial_\epsilon f_{0\pm}) \displaystyle\frac{1}{2\pi K} 
  \times \\
  \nonumber && \hskip-1.8cm
  \left[\cos\theta\frac{\cos\phi}{1+\alpha^2\pm 2\alpha\sin\phi} +
        \sin\theta\frac{1+\alpha^2}{1-\alpha^2}
        \frac{\pm \alpha+\sin\phi}{1+\alpha^2\pm 2\alpha\sin\phi}
        -\right.\\
  && \hskip3cm\left. \nonumber
        -\sin\theta\frac{\pm \alpha}{1-\alpha^2}\right]
\end{eqnarray}
for $|\alpha|<1$. 
The $1/\tau$ approach (see Appendix~\ref{sec-AppC}) leads to a
similar but not identical approximative result
\begin{eqnarray}\label{eq-33}
  f_\pm(\phi,\theta) &=& f_0 - e{\cal E}v 
  (-\partial_\epsilon f_{0\pm}) \displaystyle\frac{1}{2\pi K} 
  \times \\
  \nonumber && \hskip-1.5cm
  \left[\cos\theta\frac{\cos\phi}{1+\alpha^2\pm 2\alpha\sin\phi} +
        \sin\theta\frac{\sin\phi}{1+\alpha^2\pm 2\alpha\sin\phi}\right]\,,
\end{eqnarray}
and the $1/\tau^\parallel\,\&\,1/\tau^\perp$ approach gives precisely the
same result as Eq.~(\ref{eq-33}) because $1/\tau^\perp$ vanishes in the
two-band case (see Eq.~(\ref{eq-45}) in Appendix~\ref{sec-AppD}).

Remarkably, the AMR calculated from the distribution functions of
Eqs.~(\ref{eq-32}) or~(\ref{eq-31}) and of Eq.~(\ref{eq-33}), i.e. the exact
and the two approximative results, comes out to be the same 
\begin{equation}\label{eq-34}
  \mbox{AMR} = \alpha^2\,,\ |\alpha|<1\qquad 
  \mbox{AMR} = \frac{1}{\alpha^2}\,,\ |\alpha|>1\,.
\end{equation}
Plot of this function is shown in Fig.~\ref{fig-01}(b). 

For the two-band Rashba model with small $\lambda$, we thus conclude that the
discrepancy between the exact and approximative 
approaches remains only on the level of the complete
non-equilibrium distributions (in the higher order terms that do not
contribute to the current). We speculate that the equal results for AMR
were not obtained by coincidence but because the $\lambda\to 0$ system 
has a higher symmetry than the single-band model
which is chiral. These symmetries are briefly commented in
Appendix~\ref{sec-AppD}.

\section{Discussion and conclusion}
\label{sec-IV}

Let us start this discussion section 
with a remark on results shown in Fig.~\ref{fig-01}(a). The
AMR takes on a singular value of $1$ at $\alpha=1$ in all three
approaches. This reflects the $(1-\alpha^2)^{-1}$ divergences of all 
non-equilibrium distribution functions (see Eq.~(\ref{eq-24a}) for
example). The origin of this divergence is as follows: 
the scattering operator $\hat{V}$ in Eq.~(\ref{eq-07}) with $\alpha=1$ 
can annihilate one particular state $|\vec{k}+\rangle$ on the Fermi surface,
as seen from Eq.~(\ref{eq-35}) and the spin textures in Fig.~\ref{fig-02}.
The state has its spin aligned parallel to the moment of the magnetic
impurities, i.e., along the $\hat{x}$-axis. For the Rashba model this implies
that the $\vec{k}$-vector of this state is parallel to the $\hat{y}$-axis,
more precisely $\phi=-\pi/2$. Within the $1/\tau$ approach, the fact that 
$(\alpha+\sigma_x)|\vec{k}+\rangle =0$ then implies that this state has an
infinite (transport)  relaxation time as dictated by Eqs.~(\ref{eq-37})
and~(\ref{eq-40}) of the Appendix. Consequent calculation focusing also on
other states $|\vec{k}\rangle$ contributing to the current 
shows that this singularity is strong enough to produce
$\sigma_{yy}\to\infty$ when $|\alpha|\to 1$. 

The current (and AMR) calculated for $|\alpha|$ close to 1
are clearly inconsistent with the linear-response basis of our theory approach
(incorporated in Eq.~(\ref{eq-01})) and are therefore not physically relevant.
On the other hand, the impurity
operator~(\ref{eq-07}) is idealised compared to realistic systems where the
electric and magnetic part of $\hat{V}$ will depend at least slightly
differently on $\vec{k}$. This modification suffices to remove the singularity
in conductivity.

Pointing our attention more towards experiments, let us now discuss the
relevance of the Rashba model with dilute charged magnetic scatterers. Our
original motivation comes from the study of the diluted magnetic
semiconductor\cite{Jungwirth:2006_a} (Ga,Mn)As. Mn atoms, when
substituting for Ga, introduce both the magnetic moments and holes to the
material. The former via its $d$-electrons and the latter because their
valence number is one less than that of Ga. An 'electro-magnetic
scatterer' model as defined by Eq.~(\ref{eq-07}) is therefore 
relevant to describe
the Mn atoms which constitute by far the most frequent source of scattering in
(Ga,Mn)As. Indeed, it is possible to qualitatively explain trends for AMR
in (Ga,Mn)As based on this model of scattering and by neglecting the exchange
splitting of the (Ga,Mn)As spin-orbit coupled valence
band.\cite{Rushforth:2007_a,Rushforth:2007_b} 

The Rashba model employed in this article provides arguably the simplest
unpolarized spin-orbit coupled band structure in which  
the anisotropic scatterer mechanism fully determines the AMR. 
The other two mechanisms, which are the
anisotropy of the group velocity and the anisotropy of wavefunctions 
of the spin-split spin-orbit-coupled valence band and which only
quantitatively modify the calculated AMR in (Ga,Mn)As, 
are
completely absent in this model. The simplicity of the present model relies
mostly in that it is 2D and it considers two rotationally-symmetric 
rather than six warped bands of (Ga,Mn)As. The integral equation approach can
be straightforwardly extended to (Ga,Mn)As or other three-dimensional systems
with more ($n>2$) bands. However, the calculational complexity
will be considerably higher; the two functions $a_{\pm}(\phi)$ of one variable
will be replaced by $n$ functions of two variables (two angles parametrizing
the Fermi surface in three dimensions).

Turning attention towards possible experiments,
the calculations presented in this article are most relevant to asymmetric
$n$-type heterostructures doped with magnetic donors.\cite{Masek:2006_b} By
changing the Fermi level via doping, the effective strength $\alpha$ of the
electric part of the scatterer should change because the scattering amplitudes
depend on the Fermi wavevector which is a typical measure for involved
momentum transfers.\cite{Rushforth:2007_b} Consequently, by polarizing 
the magnetic moments in-plane, the AMR defined in Eq.~(\ref{eq-19}) should be
measurable and follow predictions shown in Fig.~\ref{fig-01}(a). 

An alternative to doping by magnetic donors is to use an $n$-type
heterostructure co-doped with magnetic impurities. Experimental study of a
III-V or II-VI heterostructure with dilute Mn doping and heavy
remote $n$-doping could be revealing.  
Depending on the magnetic impurity character (either acceptor or neutral), 
by varying the Fermi level, we could again effectively change $\alpha$
and/or interpolate between the single-band case ($\epsilon_F=0$) and the
two-band case ($\epsilon_F\gg \lambda k_F$). 
The challenge in this experiment would be to keep the scattering on Mn
the dominant (or at least strong) mechanism of relaxation.

Rather than these experimental suggestions, however, 
the main message of this paper
should be of theoretical character. We have presented a framework to calculate
exactly the conductivity in anisotropic systems within the semiclassical
linear-response theory. This
procedure was demonstrated on three simple and analytically solvable
models. We found that in some special cases of high symmetry the previously
employed approximate approaches may yield the same AMR as our exact theory.
In general, however, only the exact non-equilibrium solution to Boltzmann
equation of the form of
an integral equation over the whole Fermi surface, rather than 
of effective scattering rates at each individual
$\vec{k}$-point individually, provides a reliable account of the anisotropic
transport.

\section*{Acknowledgements}

The work was funded through Pr\ae mium Academi\ae{} and contracts number
AV0Z10100521, LC510, KAN400100652, FON/06/E002 of \hbox{GA \v CR}, 
and KJB100100802 of \hbox{GA AV} of the Czech republic, by ONR
through grant number onr-n000140610122, by NSF under grant number
DMR--0547875 and by the NAMASTE project (FP7 grant No. 214499). 
It is our pleasure to thank Maxim Trushin for critical comments 
and for providing us some of his unpublished and copyrighted 
calculations, Roman Grill for fruitful discussions and Vil\'em \v R\'\i{}ha
for his help with numerical checks of the presented results.

\setcounter{section}{0}

\section*{APPENDIX}

\subsection{Scattering rates}
\label{sec-AppA}

We evaluate the scattering rates using the Fermi golden rule. Probability
$w_{fi}$ of transition between states $|i\rangle$ and $|f\rangle$, induced by a
perturbation described by time-independent operator $\hat{V}$, equals
\begin{equation}\label{eq-36}
  w_{fi} = \frac{2\pi}{\hbar} |\langle f|\hat{V}| i\rangle|^2
           \delta\big(\epsilon_f - \epsilon_i\big)\,,
\end{equation}
where $\epsilon_{f/i}$ is the energy of the final/initial state.

Considering many scatterers described by the operator $\hat{V}$ distributed
randomly with areal 
density $n_i$, the scattering rate per unit reciprocal space 
between the $\vec{k}$-- and $\vec{k}'$--state  equals
\begin{equation}\label{eq-37}
  w(\vec{k},\vec{k}') = 
  \frac{2\pi}{\hbar} n_iV_0^2|\langle \vec{k}'|\hat{V}/V_0| \vec{k}\rangle|^2
           \delta\big(\epsilon_{\vec{k}} - \epsilon_{\vec{k}'}\big)
\end{equation}
within the lowest order of the Born approximation; contrary to the case of the
anomalous Hall
effect\cite{Sinitsyn:2007_a,Kovalev:2008_a}, 
this order of the Born approximation is sufficient for the
calculation of the AMR. 
Note that the dimension of the scatterer strength 
$V_0$ is J$\,\mbox{m}^2$, making $\hat{V}/V_0$
dimensionless in the Fourier space. 

Finally, assuming isotropic 
parabolic dispersion $\epsilon_{\vec{k}}=\hbar^2 k^2/2m$,
the density of states equals $m/(h\hbar)$ per spin, so that
\begin{eqnarray}\label{eq-38}
  w(\phi,\phi')&=& \displaystyle
  \frac{1}{(2\pi)^2}\int_0^\infty k'\, dk' w(\vec{k},\vec{k}')=\\ \nonumber
  && \hskip-2cm =\frac{2\pi}{\hbar} n_i V_0^2 \frac{m}{(2\pi \hbar)^2} 
  |\langle \vec{k}'|\hat{V}/V_0| \vec{k}\rangle|^2 \equiv 
  K |\langle \vec{k}'|\hat{V}/V_0| \vec{k}\rangle|^2\,.
\end{eqnarray}
This is the definition of the dimensionful constant $K$ used in
Eq.~(\ref{eq-08}) and later on. Its value determines the absolute value of
conductivity but it cancels out in the definition of the AMR, see
Eq.~(\ref{eq-19}).

\subsection{Scattering matrix elements}
\label{sec-AppB}

We calculate the matrix elements of the scattering operator $\hat{V}$ in
Eq.~(\ref{eq-07}) with respect to the basis
\begin{equation}\label{eq-35}
   |\vec{k}\pm \rangle = \frac{1}{\sqrt{2}}\left(\begin{array}{c}
1 \\ 
\mp ie^{i\phi}\end{array}\right) \frac{1}{\sqrt{A}}e^{i\vec{k}\cdot\vec{r}}
\end{equation}
where $\vec{k}=k(\cos\phi,\sin\phi)$ and $A$ is the system area. Vectors
$|\vec{k}+\rangle$ and $|\vec{k}-\rangle$ are the eigenstates of
Hamiltonian~(\ref{eq-06}) with eigenvalues $\hbar^2 k^2/2m - \lambda
|\vec{k}|$ and $\hbar^2 k^2/2m + \lambda |\vec{k}|$; 
their (expectation value of) spin $\vec{\sigma}=(\sigma_x,\sigma_y)$ is
illustrated in Fig.~\ref{fig-02}. The scattering operator $\hat{V}$ in
Eq.~(\ref{eq-07}) is expressed in the basis of plane waves times spin up and
spin down states. It does not depend on $\vec{k}$, $\vec{k}'$ so that 
it corresponds to short-range impurities ($\delta$-scatterers). 
For $a\to\infty$, this
would be a non-magnetic charged impurity of strength $aV_0$, and for $a=0$ it
is a purely magnetic impurity of strength $V_0$.

Owing to the $\delta$-scatterer character of $\hat{V}$,
the matrix elements of $\hat{V}/V_0$ in the basis~(\ref{eq-35}) 
depend on $\vec{k}$ only through
$\phi$ and not through $k=|\vec{k}|$. We take $k=k'$ and get
\begin{eqnarray*}
  \!\!\langle \vec{k}'{+}|\alpha+\sigma_x|\vec{k}+\rangle &=&
  \frac{1}{2}\big[-ie^{i\phi} + ie^{-i\phi'}+\alpha(1+e^{i(\phi-\phi')})\big]\\
  \langle \vec{k}'{-}|\alpha+\sigma_x|\vec{k}-\rangle &=&
  \frac{1}{2}\big[ ie^{i\phi} - ie^{-i\phi'}+\alpha(1+e^{i(\phi-\phi')})\big]\\
  \langle \vec{k}'{+}|\alpha+\sigma_x|\vec{k}-\rangle &=&
  \frac{1}{2}\big[ ie^{i\phi} + ie^{-i\phi'}+\alpha(1-e^{i(\phi-\phi')})\big]\\
  &=& \overline{\langle \vec{k}{-}|\alpha+\sigma_x|\vec{k}'+\rangle}\,.
\end{eqnarray*}
Taking the absolute values squared leads using Eq.~(\ref{eq-38}) 
to Eq.~(\ref{eq-29}), to Eq.~(\ref{eq-20}) (the '++' element), 
and to Eq.~(\ref{eq-08}) ('++' element with $\alpha=0$).

\subsection{The $1/\tau$ approach}
\label{sec-AppC}

Non-equilibrium distribution function in an isotropic
($\epsilon_{\vec{k}}=\epsilon_k$, and isotropic scatterer) two-band system can
be shown to be
\begin{equation}\label{eq-39}
  f_\pm(\phi,\theta) = f_\pm(\phi-\theta) = f_0 - e\vec{v}_\pm\cdot\vec{\cal E}
  \left(-\frac{\partial f_{0\pm}}{\partial\epsilon}\right)\tau_\pm\,,
\end{equation}
where the relaxation times for $+$ and $-$ bands 
may depend on $\vec{k}$ only through energy $\epsilon_{k}$. 
This fact, that
for fixed energy the relaxation time as defined in Eq.~(\ref{eq-15}) 
is constant, is a direct consequence of the scatterer isotropy
$w(\vec{k},\vec{k}')=w(\vartheta_{\vec{k}\vec{k}'})$. For clarity, we stress 
that in an $n$--band system there are the total of $n^2$ scattering rates
between pairs of bands,
\begin{equation}\label{eq-40}
  \frac{1}{\tau_{ba}} =
  \int \frac{d^2 k'}{(2\pi)^2} w_{ba}(\vec{k},\vec{k}') 
  \left[1-\frac{|\vec{v}_b(\vec{k}')|}{|\vec{v}_a(\vec{k})|}
  \cos\vartheta_{\vec{v}\vec{v}'}\right]\,,
\end{equation}
that combine into $n$ scattering times $\tau_a$, 
one for each band, according to the Matthiessen's rule\cite{Ashcroft:1976_a}
\begin{equation}\label{eq-41}
  \frac{1}{\tau_a} = \sum_b \frac{1}{\tau_{ba}}\,.
\end{equation}
We note that $\vartheta_{\vec{v}\vec{v}'}$ measures the angle between
$\vec{v}_b(\vec{k}')$ and $\vec{v}_a(\vec{k})$ but given the isotropy of 
the band structure, $\vec{v}(\vec{k})$ and $\vec{k}$ are parallel so that
$\vartheta_{\vec{v}\vec{v}'}=\vartheta_{\vec{k}\vec{k}'}$. 
Equation~(\ref{eq-15}) is a single-band variant of Eq.~(\ref{eq-40}) for
isotropic systems where $v$ drops out. 

In the $1/\tau$ approach, we simply evaluate Eq.~(\ref{eq-40}) for
$w(\vec{k},\vec{k}')\not= w(\vartheta_{\vec{k}\vec{k}'})$ and 
obtain $\vec{k}$--dependent
$1/\tau$. This is then inserted into the distribution function~(\ref{eq-39}),
losing thereby its property $f_\pm(\phi,\theta)=f_\pm(\phi-\theta)$.  

For the Rashba model with $\lambda\to 0$ we get using Eq.~(\ref{eq-38}) and
Appendix~\ref{sec-AppB} the following
\begin{equation}\label{eq-43}\begin{array}{rcl}
\tau^{-1}_{(++/--)}(\phi)\!\!&=&\!\! 
   K\pi \big((+/-)2\alpha\sin\phi+3-2\sin^2\phi+\alpha^2\big)
  \\ 
\tau^{-1}_{(+-/-+)}(\phi)\!\!&=&\!\! 
   K\pi \big((-/+)6\alpha\sin\phi+1+2\sin^2\phi+3\alpha^2\big)
  \,.
  \end{array}
\end{equation}
The scattering rates in the two-band model are thus as simple as
\begin{equation}\label{eq-42}
  \frac{1}{\tau_\pm(\phi)} = \frac{1}{\tau_{+\pm}(\phi)}+
                             \frac{1}{\tau_{-\pm}(\phi)} =
  \frac{1/(4\pi K)}{1\pm 2\alpha\sin\phi + \alpha^2}\,.
\end{equation}
This result, plugged into Eq.~(\ref{eq-39}), produces the non-equilibrium
distribution function~(\ref{eq-33}) within the $1/\tau$ approach.

The relaxation time for the single band model is simply
$1/\tau(\phi)=1/\tau_{++}(\phi)$. Setting here $\alpha=0$ leads via
Eq.~(\ref{eq-39}) to Eq.~(\ref{eq-17}).

\subsection{The $1/\tau^\parallel\,\&\,1/\tau^\perp$ approach}
\label{sec-AppD}

The prescription for the non-equilibrium distribution function suggested by
Schliemann and Loss\cite{Schliemann:2003_b} can be summarized as follows: (a)
evaluate the 'standard' formulae~(\ref{eq-41},\ref{eq-40}) and denote the 
result as
$1/\tau^\parallel_a(\phi)$; (b) calculate $1/\tau^\perp_a(\phi)$ using
formulae identical to Eqs.~(\ref{eq-40},\ref{eq-41}) save the replacement of
the bracket in Eq.~(\ref{eq-40}) 
by $|v_b(\vec{k}')|/|v_a(\vec{k})|\,\sin\vartheta_{\vec{v}\vec{v}'})$; 
(c) write down the distribution function as
\begin{eqnarray}\label{eq-44}
  f_\pm(\phi,\theta) &=& f_0 - e|\vec{v}_\pm|\,|\vec{\cal E}|
  (-\partial_\epsilon f_{0\pm})\times \\
  \nonumber && \hskip-2cm\tau_\pm^\parallel \! \left[
  \cos(\phi-\theta)
  \frac{(\tau_\pm^\perp)^2}{(\tau_\pm^\parallel)^2+(\tau_\pm^\perp)^2}\!
  +\! \sin(\phi-\theta)
  \frac{\tau_\pm^\perp \tau_\pm^\parallel}%
{(\tau_\pm^\parallel)^2+(\tau_\pm^\perp)^2}\!  
  \right]\,.
\end{eqnarray}

Several remarks are in order. (i) Whenever $1/\tau_\pm^\perp$ vanishes,
Eq.~(\ref{eq-44}) simplifies to Eq.~(\ref{eq-39}) of the $1/\tau$ approach.
(ii) This $1/\tau^\parallel\,\&\,1/\tau^\perp$ approach is suitable for the
description of isotropic scatterers (the amplitude depends only on the angle
between $\vec{k}$ and $\vec{k}'$, the incoming and outgoing wave) which
however may exhibit an asymmetry (or better chirality), i.e. scatter more
clockwise than counterclockwise --- such as it is the case with skew
scattering in the anomalous Hall effect. (iii) Contrary to the statement of
Ref.~\onlinecite{Schliemann:2003_b}, the distribution function~(\ref{eq-44})
is not the exact solution to Eq.~(\ref{eq-26}) for a general anisotropic
system. The derivation of Eq.~(\ref{eq-44}) presented in
Ref.~\onlinecite{Schliemann:2003_b} is only valid if the expressions
$$
\frac{\tau_a^\parallel}{1+\big(\tau_a^\parallel/\tau_a^\perp\big)^2}\,,
  \quad
\frac{\tau_a^\perp}{1+\big(\tau_a^\perp/\tau_a^\parallel\big)^2}\,,  
$$
given by Eq.~(27,28) of that reference are constant for each band
(i.e. $\phi$--independent in our case). The most general distribution function
this approach can therefore correctly capture must have the form
$$
  f(\phi,\theta) - f_0 = C_1\cos(\phi-\theta) + C_2\sin(\phi-\theta)
$$
while as the examples in Section~\ref{sec-II} show, the non-equilibrium
distribution can have finer details than those of period $2\pi$ in the angular
variable $\phi$ (and these details, when completely neglected, may even
lead to wrong values of the constants $C_1$, $C_2$ above). This original
neglect of Ref.~\onlinecite{Schliemann:2003_b} was later corrected by one of
its authors\cite{Trushin:2006_a} in the context of the specific Hamiltonian
considered.\cite{Schliemann:2003_b} However, a general procedure for exact
solution of the Boltzmann equation was not given.

In our specific model, as described by the scattering matrix elements of
Appendix~\ref{sec-AppB}, we get
\begin{equation}\label{eq-46}\begin{array}{rcl}
[\tau^\perp_{(++/--)}(\phi)]^{-1} &=&K\pi\cos\phi\big((+/-)\alpha+\sin\phi\big)
\\[0mm]
 [\tau^\perp_{(+-/-+)}(\phi)]^{-1}&=&
   K\pi\cos\phi\big((+/-)\alpha-\sin\phi\big)\,.
\end{array}
\end{equation}
For the two-band model,
\begin{equation}\label{eq-45}
  \frac{1}{\tau^\perp_+(\phi)} = \frac{1}{\tau^\perp_{++}(\phi)}
+\frac{1}{\tau^\perp_{-+}(\phi)} =0\,,\qquad \frac{1}{\tau^\perp_-(\phi)}=0\,,
\end{equation}
so that the $1/\tau^\parallel\,\&\,1/\tau^\perp$ approach reduces to the
$1/\tau$ approach in line with the comment after Eq.~(\ref{eq-44}).
The single-band case, however, has a finite $\tau^\perp\equiv \tau_{++}^\perp$
so that the two approaches give different results. This is not surprising,
since each Rashba band has a chiral spin texture but both of them together 
form a non-chiral pair, provided they have both the same Fermi $k$ (as it
happens for $\lambda\to 0$), see Fig.~\ref{fig-02}. The asymmetry of
scattering expressed by $1/\tau^\perp$ thus vanishes in our two-band model.

To obtain the non-equilibrium distribution~(\ref{eq-18})
within the $1/\tau^\parallel\,\&\,1/\tau^\perp$ approach, we have to take
$1/\tau^\parallel_{++}$ of Eq.~(\ref{eq-43}), $1/\tau^\perp_{++}$ of
Eq.~(\ref{eq-46}), insert them into Eq.~(\ref{eq-44}) and expand
$\cos(\phi-\theta)$, $\sin(\phi-\theta)$ in terms of $\cos\theta$ and
$\sin\theta$.

\subsection{Partitioning method}
\label{sec-AppE}

The actual infinite system of linear equations for variables
$a_0, a_{c1}, a_{s1}, \ldots$
appropriate for the {\em single-band model} assumes a structure suitable for
partitioning if we perform the coordinate transformation 
$\ihp = \pi/2 - \phi$. We will now solve 
the integral equation~(\ref{eq-05a}) using this
coordinate (and use $\ihp$ throughout
Appendices~\ref{sec-AppE}~and~\ref{sec-AppF}) and transform the result back
before we use it in Eq.~(\ref{eq-21}).

The equation to be solved is now
\begin{equation}\label{eq-53}
  \bar{w}(\ihp)\atr (\ihp) - \int d\ihp' w(\ihp,\ihp')\atr (\ihp') = \sin\ihp
\end{equation}
with
\begin{eqnarray*}
  w(\ihp,\ihp') &=& \textstyle\frac{1}{2} K \big[ 1 + \cos(\ihp+\ihp') +
\alpha^2\big(1+\cos(\ihp-\ihp')\big)] +\\
 &&  + \alpha(\cos\ihp + \cos\ihp') \\
     \bar{w}(\ihp)&=& \pi K (1+\alpha^2+2\alpha\cos\ihp)\,.
\end{eqnarray*}
Inserting
$$
  \atr (\ihp) = \atr_0 + \atr_{c1} \cos\ihp + \ldots + \atr_{s1}\sin\ihp+
  \ldots
$$
into Eq.~(\ref{eq-53}), and comparing the coefficients at the constant,
$\cos\ihp$, $\sin\ihp$, $\cos 2\ihp$, $\cos 3\ihp$, 
$\ldots$, $\sin 2\ihp$, $\ldots$ terms,
we obtain the following infinite system of linear
equations for $\atr_0, \atr_{c1}, \atr_{s1}, \atr_{c2}, \atr_{c3},
\ldots, \atr_{s2}, \atr_{s3},\ldots$ (in this order):
\def\aa{\alpha}
\begin{equation}\label{eq-47}
  \hskip-4mm\left(\begin{array}{ccc|cccc|cccc||c} 
  * & * & * &    0 & 0 & 0 & \ldots & 0 & 0 & 0 & \ldots & * \\[0cm]
  * & * & * &    \aa & 0 & 0 & \ldots & 0 & 0 & 0 & \ldots & * \\[0cm]
  * & * & * &    0 & 0 & 0 & \ldots & \aa & 0 & 0 & \ldots & * \\ \hline
  0 & \aa & 0 &\!1+\aa^2\!\!\!&\aa & 0 & \ldots & 0 & 0 & 0 & \ldots & 0 \\
  0 & 0 & 0 & \aa &\!\!\!1+\aa^2\!\!\!&\aa &      & 0 & 0 & 0 & \ldots & 0 \\
  0 & 0 & 0 & 0 & \aa &\!\!\!1+\aa^2\!\!\!&     & 0 & 0 & 0 & \ldots & 0 \\
  \vdots & &&\vdots& &&\ddots       &\vdots & & & \vdots & \vdots \\ \hline
  0 & 0 & \aa & 0    & 0 & 0 &\!\ldots & 1+\aa^2 & \aa& 0&\ldots& 0 \\[-2mm]
  0 & 0 & 0 & 0    & 0 & 0 & \ldots & \aa & \ddots & &\vdots& 0 \\
  \vdots &&&&&&&&&&\vdots &\vdots
    \end{array}\right)
\end{equation}
The double line separates the left
and right-hand side of the equations.
The twelve asterisks in the first three lines of the system~(\ref{eq-47}) 
correspond to the $3\times 3$ system~(\ref{eq-21}), and the value of these
coefficients will be unimportant within this Appendix.

It is apparent that the system~(\ref{eq-47}) is almost block-diagonal. The
partitioning method takes advantage of this structure and aims at solving
three independent systems corresponding to groups 
$\{\atr_0,\atr_{c1},\atr_{s1}\}$, $\{\atr_{c2},\atr_{c3},\ldots\}$, and
$\{\atr_{s2},\atr_{s3},\ldots\}$ of the original variables. The basic idea 
is to treat the only non-zero element of the off-diagonal block as a
right-hand-side term. In explicite terms, we rewrite for example 
the fourth and fifth equations of the system~(\ref{eq-47})
\begin{eqnarray*}
[ \alpha\atr_{c1}+(1+\alpha^2)\atr_{c2}+\alpha\atr_{c3}]\cos 2\ihp &=& 0\\[0cm]
[ \alpha\atr_{c2}+(1+\alpha^2)\atr_{c3}+\alpha\atr_{c4}]\cos 3\ihp &=& 0
\end{eqnarray*}
as
\begin{eqnarray*}
  (1+\alpha^2)\atr_{c2}+\alpha\atr_{c3}\phantom{\alpha\atr_{c2}+}\      
          &=&-\alpha\atr_{c1} \equiv\Delta\\[0cm]
  \alpha\atr_{c2}+(1+\alpha^2)\atr_{c3}+\alpha\atr_{c4} &=& 0\,.
\end{eqnarray*}
The system of all 'cosine-term' equations of the system~(\ref{eq-47})
(starting with $\cos 2\ihp$) can now be solved as a function of $\Delta$. In
other words, we are treating the central block of the
matrix~(\ref{eq-47}). The still-infinite system to be solved is
\begin{equation}\label{eq-48}
\left(\begin{array}{ccccc||c}
1+\alpha^2 & \alpha & 0 & 0 & \ldots & \Delta \\
\alpha & 1+\alpha^2 & \alpha & 0 & \ldots & 0 \\
0 & \alpha & 1+\alpha^2 & \alpha & \ldots & 0 \\
\vdots &  &   & \ddots & \vdots & \vdots 
\end{array}\right)\,.
\end{equation}
For the purposes of solving later 
the $3\times 3$ system~(\ref{eq-21}), we in fact need to
know only a part of the solution, namely $\atr_{c2}$. 
To this end, linear algebra
gives us a very quick answer. If we denote by $D$ the determinant of the
infinite matrix left from the double line in~(\ref{eq-48}), and by $D_n$ the
determinant of the analogous $n\times n$ matrix, then if the
system~(\ref{eq-48}) were finite,
$$
  \atr_{c2} = \frac{\Delta D_{n-1}}{D_n}\,,
$$
where the numerator equals the determinant of the $n\times n$ matrix left from
the double line of~(\ref{eq-48}) with first column replaced by the column
right from the double line. Considering $n\to\infty$, we immediatelly (after
transformation $a_{c2}=-\atr_{c2}$) get
$\atr_{c2}=-\Delta$ as given in the first line of Eq.~(\ref{eq-22}).
This answer is, however, not completely correct.

The caveat of this procedure is that we should have been careful about taking
the limit $D=\lim_{n\to\infty} D_n$. 
It turns out that the limit is finite only for
$|\alpha|<1$ and then $D=1/(1-\alpha^2)$ so that only in this case $\lim
D_{n-1}/D_n=(\lim D_{n-1})/(\lim D_{n})$ which is obviously equal to one. The
determinant $D$ is infinite for $|\alpha|>1$ and only $\lim_{n\to\infty}
D_{n-1}/D_n$ remains finite, namely equal 
to $1/\alpha^2$ as one can readily see from the
explicit formula
$$
  D_n=1+\alpha^2 + \alpha^4 + \ldots + \alpha^{2n}\,.
$$
In conclusion, we find
\begin{equation}\label{eq-49}
  \atr_{c2}=\left\{ \begin{array}{ll} \Delta     & \mbox{ for }|\alpha|<1 \\
                                      \Delta/\alpha^2 & \mbox{ for }|\alpha|>1
    \end{array}\right.
\end{equation}
and the transformation back from $\ihp$ to $\phi$ implies $a_{c2}=-\atr_{c2}$
and $\atr_{c1}=a_{s1}$.

Literally the same procedure works for the 'sine-term' equations of the
system~(\ref{eq-47}), i.e. the lower-right block. The only difference is now
that $\Delta=-a\atr_{s1}$ and we use $a_{s2}=\atr_{s2}$, $\atr_{s1}=a_{c1}$. 
These two results, $a_{c2},a_{s2}$ with the corresponding definitions of
$\Delta$ are summarized as Eq.~(\ref{eq-22}). 

The key feature needed for this partitioning method is that $\Delta$ is a
function only of $\atr_{c1}$ ($\atr_{s1}$) and not of higher-order
coefficients like $\atr_{c3}$. In this way, the $3\times 3$ system of
equations~(\ref{eq-21}) becomes closed after $a_{c2}$ and $a_{s2}$ have been
inserted.

Finally, we stress, that if the coupling between the three subsystems had been
neglected from the very beginning --- this amounts to setting to zero the four
elements in the off-diagonal blocks in the system~(\ref{eq-47}) --- the
solution of the $3\times 3$ subsystem represented by the asterisks would have
been different. In this way, even though $\cos 2\phi$ and other higher terms
do not contribute to the current calculated from the non-equilibrium
distribution~(\ref{eq-03}), their complete neglect from the beginning may
produce wrong coefficients in the $\cos\phi$ and $\sin\phi$ terms.

\subsection{Partitioning method -- two bands}
\label{sec-AppF}

In the case of two bands, 
we obtain two infinite systems of linear equations identical to
the system~(\ref{eq-47}), one for variables with '+' index, another for those
with '$-$' index, see Eq.~(\ref{eq-30}). Although the two systems are now
coupled, the direct coupling exists only via variables $\atr_{0\pm}$,
$\atr_{c1\pm}$, $\atr_{s1\pm}$ corresponding to the upper left block. The
partitioning method can therefore be independently carried out in the '$+$'
and '$-$' sector.

For all four infinite subsystems, of which the system~(\ref{eq-48}) is one, we
obtain the almost the same result
\begin{equation}\label{eq-50}
  \atr_{cn+},\, \atr_{sn+}
  =\left\{ \begin{array}{ll} \Delta (-\alpha)^{n-2} & \mbox{ for }|\alpha|<1 \\
                             \Delta/(-\alpha)^{n} & \mbox{ for }|\alpha|>1\,,
    \end{array}\right.
\end{equation}
for $n\ge 2$ and with appropriate definition of $\Delta$ for each subsystem,
while $\atr_{cn-},\atr_{sn-}$ obey Eq.~(\ref{eq-50}) with $-a$ replaced by $a$.
All coefficients in
the series~(\ref{eq-30}) are thus non-zero. Nevertheless, Eq.~(\ref{eq-30})
can still be summed up using
$$
  \sum_{n=0}^\infty (-\alpha)^n \cos n\ihp = 
  \frac{1+\alpha\cos\ihp}{1+\alpha^2+2\alpha\cos\ihp}
$$
and a similar formula for sines. We now transform back from $\ihp$ to $\phi$,
use $\Delta=\mp \alpha\atr_{c1\pm}$ for the cosine $\pm$ parts of
Eq.~(\ref{eq-30}) and $\Delta=\mp \alpha\atr_{s1\pm}$ for its sine parts,
transform back $\atr_{c1\pm}=a_{s1\pm}$, $\atr_{s1\pm}=a_{c1\pm}$ and finally
get 
\begin{equation}\label{eq-51}\begin{array}{rcl}
  a_+(\phi)&=&a_{0+} + \\ && \hskip-.8cm\displaystyle
   a_{c1+}\frac{\cos\phi}{1+\alpha^2+2\alpha\sin\phi} +
   a_{s1+}\frac{\alpha+\sin\phi}{1+\alpha^2+2\alpha\sin\phi}  \\
  a_-(\phi)&=&a_{0+} + \\ && \hskip-.8cm\displaystyle
   a_{c1-}\frac{\cos\phi}{1+\alpha^2-2\alpha\sin\phi} +
   a_{s1-}\frac{-\alpha+\sin\phi}{1+\alpha^2-2\alpha\sin\phi}\,.
  \end{array}
\end{equation}
Plugging the values of $a_{0+},a_{c1+},a_{s1+}$ from Eq.~(\ref{eq-52}) 
into Eq.~(\ref{eq-51}), repeating an analogous procedure for the $b$'s in
Eq.~(\ref{eq-28b}) and inserting the results into Eq.~(\ref{eq-27}), 
we arrive at Eq.~(\ref{eq-32}).

\subsection{Boltzmann equation in general 2D anisotropic systems}
\label{sec-AppG}

Results of Section~\ref{sec-II} were derived for a special class of 2D systems
where the band structure remains isotropic and the anisotropy is only
introduced through the scatterer and the scattering rate
$w(\vec{k},\vec{k}')$. 

The results of Eqs.~(\ref{eq-05a},\ref{eq-05b}) for single-band or of
Eqs.~(\ref{eq-28a},\ref{eq-28b}) for two-band system have to be slightly
modified  for anisotropic
2D band structure. The wavevectors $\vec{k}, \vec{k}'$ of Eq.~(\ref{eq-01}) or 
Eqs.~(\ref{eq-26}) will still be bound to the Fermi level 
$\epsilon_F$ but their magnitude now depends on $\phi$. That is, we have
$k=k(\phi)$ and $\vec{k}=k\hat{n} = k(\cos\phi,\sin\phi)$.
We also tacitly assume that in each band and for each $\hat{n}$ there is only
one solution $k$ to $\epsilon_{k \hat{n}} = \epsilon_F$. The
calculation of $w(\phi,\phi')$, compared to what 
is done in Appendix~\ref{sec-AppA}, becomes 
\begin{eqnarray}\label{eq-54}
  w(\phi,\phi') &=& \displaystyle \frac{1}{(2\pi)^2} \int_0^\infty k'\, dk'
  w(\vec{k},\vec{k}') = \\ \nonumber
  &=&\frac{2\pi}{\hbar} n_i V_0^2 
  \left|\nabla_{\vec{k}'}\epsilon_{\vec{k}'}\cdot \vec{k}'/{k'}^2\right|^{-1}\!
  |\langle\vec{k}'|\hat{V}/V_0|\vec{k}\rangle|^2\,. 
\end{eqnarray}
The last expression should be understood as a function of $\phi$,
$\phi'$ only; the derivative and $\vec{k},\vec{k}'$ are to be taken at the
Fermi level, so that e.g. 
$\vec{k}'=k'(\cos\phi',\sin\phi')$ and $\epsilon_{\vec{k}'}=\epsilon_F$.  

Further, the expression $\vec{\cal E}\cdot \vec{v}(\vec{k})$ 
in Eq.~(\ref{eq-01}) is no longer
simply ${\cal E}v \cos(\theta-\phi)$. First of all, $v=v(\phi)$ and moreover
$\vec{v}$ need not be parallel with $\vec{k}$. Formally, we could replace
$\theta-\phi$ in Eq.~(\ref{eq-04}) by $\theta-\xi(\phi)$ with $\xi$ defined by
$\vec{v}(\vec{k})=(\cos\xi,\sin\xi)v(\phi)$.
Single-band equations~(\ref{eq-05a},\ref{eq-05b}) should be replaced by
\begin{eqnarray}\label{eq-55a} \hskip-5mm
  \cos\xi(\phi) &=& \bar{w}(\phi)\, a(\phi) - 
  \int d\phi' \frac{v(\phi')}{v(\phi)} w(\phi,\phi') a(\phi')\\ \label{eq-55b}
\hskip-5mm
  \sin\xi(\phi) &=& \bar{w}(\phi)\, b(\phi) - 
  \int d\phi' \frac{v(\phi')}{v(\phi)} w(\phi,\phi') b(\phi')\,.
\end{eqnarray}
These two equations for $a(\phi)$ and $b(\phi)$ 
are still completely decoupled.  With some luck,
$\cos\xi(\phi)$ can be reasonably expanded in terms of cosines and sines of
$\phi$ and higher multiples of $\phi$ but the $v(\phi)$ and $v(\phi')$ terms
will most likely make an analytical solution of Eq.~(\ref{eq-55a}) impossible
for realistic anisotropic Fermi surfaces. The solution is, however, not
difficult to obtain by numerical means. After discretization of the angular
variable $\phi$ into $n$ steps, Eq.~(\ref{eq-55a}) constitutes an $n\times n$
system of linear equations.

Once $a(\phi), b(\phi)$ are known, the non-equilibrium distribution function
is readily written as
\begin{eqnarray}\label{eq-56}
  f(\vec{k},\vec{\cal E}) - f_0 &=& f(\phi,\theta) - f_0 = \\ \nonumber
  && -e {\cal E} v(\phi) (-\partial_\epsilon f_0) 
     \big[ a(\phi)\cos\theta + b(\phi)\sin\theta\big]\,.
\end{eqnarray}
Note that the spectral function $-\partial_\epsilon f_0=\delta(\epsilon_F
-\epsilon_{\vec{k}})$ depends now both on $k$ and $\phi$.

A rather straightforward generalization of Eq.~(\ref{eq-56}) and the
appropriate pair of integral equations~(\ref{eq-55a},\ref{eq-55b}) 
to multiband systems is possible. For instance, the analogy of the two
coupled Eqs.~(\ref{eq-28a}) for anisotropic band structure reads
\begin{eqnarray*}
  \cos\xi_+(\phi) &=& \bar{w}_+(\phi)\, a_+(\phi) - \int d\phi' \times\\
  &&\displaystyle  \hskip-20mm \times
  \left[\frac{v_+(\phi')}{v_+(\phi)}w_{++}(\phi,\phi') a_+(\phi') 
   + \frac{v_-(\phi')}{v_+(\phi)}w_{+-}(\phi,\phi') a_-(\phi')\right]\\[2mm]
  \cos\xi_-(\phi) &=& \bar{w}_-(\phi)\, a_-(\phi) - \int d\phi' \times\\
  &&\displaystyle  \hskip-20mm \times
  \left[\frac{v_-(\phi')}{v_-(\phi)}w_{--}(\phi,\phi') a_-(\phi') 
            + \frac{v_+(\phi')}{v_-(\phi)}w_{-+}(\phi,\phi') a_+(\phi')\right]
\end{eqnarray*}
where $\vec{v}_\pm =(\cos\xi_\pm,\sin\xi_\pm)v_\pm(\phi)$ are the Fermi
velocities of the two bands and the four 
quantities $w_{\pm\pm}(\phi,\phi')$ have to be calculated in the spirit of
Eq.~(\ref{eq-54}).




\end{document}